\documentclass[useAMS,usenatbib,aps,prd,onecolumn,preprintnumbers,nofootinbib,amsmath,superscriptaddress]{revtex4}
\usepackage{graphicx}
\usepackage{longtable}
\usepackage{float}
\usepackage{dcolumn}
\usepackage{bm}
\usepackage{appendix}
\usepackage{multirow}

\newcommand{\Mpc}{\mathrm{Mpc}}

\begin{document}

\title{A Bayesian analysis of inflationary primordial spectrum models using Planck data}

\author{Simony Santos da Costa}
\affiliation{Departamento de Astronomia, Observat\'orio Nacional, 20921-400, 
Rio de Janeiro, RJ, Brazil}
\affiliation{Dipartimento di Fisica, Universit\`a di Napoli ``Federico II",
Compl. Univ. di Monte S. Angelo, Edificio G, Via Cinthia, I-80126, Napoli, Italy}

\author{Micol Benetti}
\affiliation{Departamento de Astronomia, Observat\'orio Nacional, 20921-400, 
Rio de Janeiro, RJ, Brazil}

\author{Jailson Alcaniz}
\affiliation{Departamento de Astronomia, Observat\'orio Nacional, 20921-400, 
Rio de Janeiro, RJ, Brazil}
\affiliation{Physics Department, McGill University, Montreal, QC, H3A 2T8, Canada}

\date{\today}

\begin{abstract}
The current available CMB data show an anomalously low value of the CMB temperature fluctuations at large angular scales ($\ell < 40$). This lack of power is not explained by the minimal $\Lambda$CDM model, and one of the possible mechanisms explored in the literature to address this problem is the presence of features in the primordial power spectrum (PPS) motivated by the early universe physics. In this paper, we analyse a set of cutoff inflationary PPS models using a Bayesian model comparison approach in light of the latest Cosmic Microwave Background (CMB) data from the Planck Collaboration. Our results show that the standard power-law parameterisation is preferred over all models considered in the analysis, which motivates the search for alternative explanations for the observed lack of power in the CMB anisotropy spectrum. 
\end{abstract}

\pacs{pacs}

\maketitle

\section{Introduction}
The predictions of the minimal cosmological constant ($\Lambda$) + cold dark matter (CDM) model with a primordial potential spectrum (PPS) of the power-law type shows a good agreement with the current Cosmic Microwave Background (CMB) observations~\cite{Ade:2015lrj, planck_cosmo_param}. 
However, despite of its consistency with CMB observations there are some tensions which emerge when different data sets at intermediate scales ($z \lesssim 1$) are analysed in the context of this model. Some examples are the present-day value of the Hubble parameter, estimates of the power spectrum amplitude on scales of $8h^{-1}$ Mpc, and measurements of the matter density parameter (see, e.g., \cite{Bull:2015stt} for a general discussion and the references therein for details).

Another intriguing  aspect of the current data are the features on the CMB temperature power spectrum which are not fully explained by the standard $\Lambda$CDM cosmology. In particular, the lack of angular power at large scale (see \cite{Schwarz:2015cma, Copi:2013cya, Yoho:2013tta, Copi:2013zja} for an exhaustive discussion) was firstly noticed by the Cosmic Background Explorer (COBE)~\cite{jing} satellite and later confirmed by the Wilkinson Microwave Anisotropy Probe (WMAP) experiment~\cite{Bennett:2012zja} and by the Planck satellite~\cite{Aghanim:2015xee}. Although the deviation from the $\Lambda$CDM best-fit prediction lies in the cosmic variance uncertainty, the possibility that it is due to a physical mechanism in the early universe cannot be excluded.
Indeed, if one admits that CMB anisotropies are sourced by quantum fluctuations generated during inflation, thus this lack of power could be explained by some mechanisms, such as a negative running of the spectral index~\cite{kosowsky, easter, kobayashi, czerny, Benetti:2013wla} or a feature at large wavelengths of the primordial power spectrum able to produce a depletion of power. Such features can be obtained, for example, from a brief violation of the slow-roll condition~\cite{Mortonson:2009qv,Hazra:2010ve,Benetti:2013cja}, 
or assuming an inflationary epoch preceded by matter or radiation domination~\cite{Vilenkin:1982wt}, 
or considering an oscillating scalar field which couples to the inflaton~\cite{Burgess:2002ub}, 
or also assuming that the onset of a slow-roll phase coincides with the time when the largest observable scales exited the Hubble radius during inflation~\cite{sinhasou}.

Since features in the PPS has been the most common mechanism to address the problem of lack of power at low multipoles in the CMB anisotropy spectrum, in what follows we present a Bayesian model selection analysis of single-field inflationary models  able to explain this problem, the so-called ``cutoff PPS models". Such models ranges from the simplest empirical models, without assuming any physical mechanism responsible for the feature~\cite{bridges1,bridges2,iqbal2015,cline, contaldi}, to most complex ones, where the modulations in the PPS are obtained by a given physical mechanism~\cite{Vilenkin:1982wt,powell,wangI, iqbal2015,sinhasou,contaldi}. Our study differs from previous investigations (see, e.g., Ref.~\cite{iqbal2015}) in two aspects. First, we use the most recent data from the Planck Collaboration. Second, we perform an accurate Bayesian analysis of such PPS models in order to investigate their compatibility with the high accuracy of the Planck data. Differently from the statistical methods used in Ref.~\cite{iqbal2015}, the Bayesian model comparison selects the best-fit model by achieving the best compromise between quality of fit and predictivity  and by evaluating whether the extra complexity of a model is required by the data, preferring the model that describes the data well over a large fraction of their prior volume.

This paper is organised as follows. Sec.~\ref{sec:model} briefly introduces the inflationary model and reviews the class of inflationary models considered in this work. In Sec.~\ref{sec:method} we discuss the observational data sets and priors used in the analysis as well as the Bayesian model selection method adopted. In Sec.~\ref{sec:results} we discuss the results and present a comparison with previous analysis. We end the paper by summarising the main results in Sec.~\ref{sec:conclusion}.

\section{Inflationary scenarios}
\label{sec:model}

The standard inflationary dynamics is governed by the action,
\begin{eqnarray}
S=\int{d^4x\sqrt{-g}\left[\frac{M_{Pl}^2}{2}R-\frac{1}{2}g^{\mu\nu}\partial_{\mu}\phi\partial_{\nu}\phi-V(\phi)\right]},
\end{eqnarray}                                                                                                                                                                                                                             
where $R$ is the four-dimensional Ricci scalar derived using the metric $g_{\mu\nu}$ and $V(\phi)$ is the potential energy of the inflaton field. The dynamics of the inflation field is governed by the Friedman and Klein-Gordon equations
\begin{eqnarray}
H^2=\frac{1}{3M_{Pl}^2}\left(\frac{1}{2}\dot{\phi}^2 + V(\phi)\right),\label{fried_eq} \\
\ddot{\phi}+3H\dot{\phi}+V'(\phi)=0,\label{phi_eq}
\end{eqnarray} 
where $H$ is the Hubble parameter and the derivatives with respect to the cosmic time and scalar field are denoted, respectively, by dots and primes.  In the slow-roll regime inflation is realised
by the single scalar field $\phi$ slowly rolling down its potential $V(\phi)$. It is characterised by the slow-roll parameters:
\begin{eqnarray}
\epsilon&\equiv&-\frac{\dot{H}}{H^2}=\frac{M_{Pl}^2}{2}\left(\frac{V'(\phi)}{V(\phi)}\right)^2,\\
\eta&\equiv&\epsilon+\delta=M_{Pl}^2\left(\frac{V''(\phi)}{V(\phi)}\right),\\
\xi&\equiv&M_{Pl}^{2}\frac{V'(\phi)V'''(\phi)}{V(\phi)^2},
\end{eqnarray}
where $\displaystyle \delta=-{\ddot{\phi}}/{H\dot{\phi}}$. In terms of these parameters, inflation happens when $\epsilon<<1$ and lasts for a sufficiently long time for $\eta<<1$. One should note that the acceleration condition ($\epsilon<<1$) also implies that the comoving Hubble radius $(aH)^{-1}$ is a decreasing function of time. 

In order to study the curvature perturbation $\mathcal{R}$ produced due to fluctuations in the scalar field $\phi$, one can use the Mukhanov-Sasaki equation~\cite{weinbergbook}
\begin{eqnarray}
u''_k+\left(k^2-\frac{z''}{z}\right)u_k=0,
\end{eqnarray}
where $u\equiv -z\mathcal{R}$ and $z\equiv a\dot{\phi}/H$. The solution of this equation can be obtained by considering a Bunch-Davies vacuum, in which all modes of cosmological interest are well inside the horizon at sufficiently early times ($k/aH\gg1$), such that~\cite{mukhanovbook}
\begin{eqnarray}
u_k(\tau)\rightarrow\frac{1}{\sqrt{2k}}e^{-ik\tau}.
\end{eqnarray}
We can define the primordial power spectrum of curvature perturbations $\mathcal{P_R}(k)$ in terms of the vacuum expectation value of $\mathcal{R}$
\begin{eqnarray}
<\mathcal{R}^{*}(k)\mathcal{R}(k')>=\frac{2\pi^2}{k^3}\delta^{3}(k-k')\mathcal{P_R}(k),
\end{eqnarray}
where $\delta$ is the Dirac delta function and the factor $\displaystyle2\pi^2/k^3$ is chosen to obey the usual Fourier conventions. On the other hand, $\mathcal{P_R}(k)$ is related to $u_k$ and $z$ via:
\begin{eqnarray}
\mathcal{P_R}(k)=\frac{k^3}{2\pi^2}\left|\frac{u_k}{z}\right| ^2.
\end{eqnarray}

The simplest shape of the primordial power spectrum in the standard $\Lambda$CDM cosmology is the power law parameterization, which can be obtained considering the slow-roll approximation of the single-inflaton field~\cite{arxiv:9504071}:
\begin{eqnarray}
\mathcal{P}(k)=A_s\left(\frac{k}{k_0}\right)^{n_s-1},
\label{eqPPSPL}
\end{eqnarray}
where $n_s$ is the spectral index, which is constant for power law models ($n_s = 1$ corresponds to a scale-invariant, Harrison-Zel'dovich-Peebles power spectrum), $A_s$ is the spectral amplitude and $k_0$ is the pivot scale set equal to $0.05$ Mpc$^{-1}$. In terms of the slow-roll parameters we can evaluate the primordial power spectrum parameters as 
\begin{eqnarray}
A_s\simeq \frac{V(\phi)}{24\pi^2\epsilon M_{Pl}^4} \quad \mbox{and} \quad n_s\simeq 1+2\eta-6\epsilon
\end{eqnarray}

%
\begin{figure}[t]
 \begin{center}
 \includegraphics[scale=0.47]{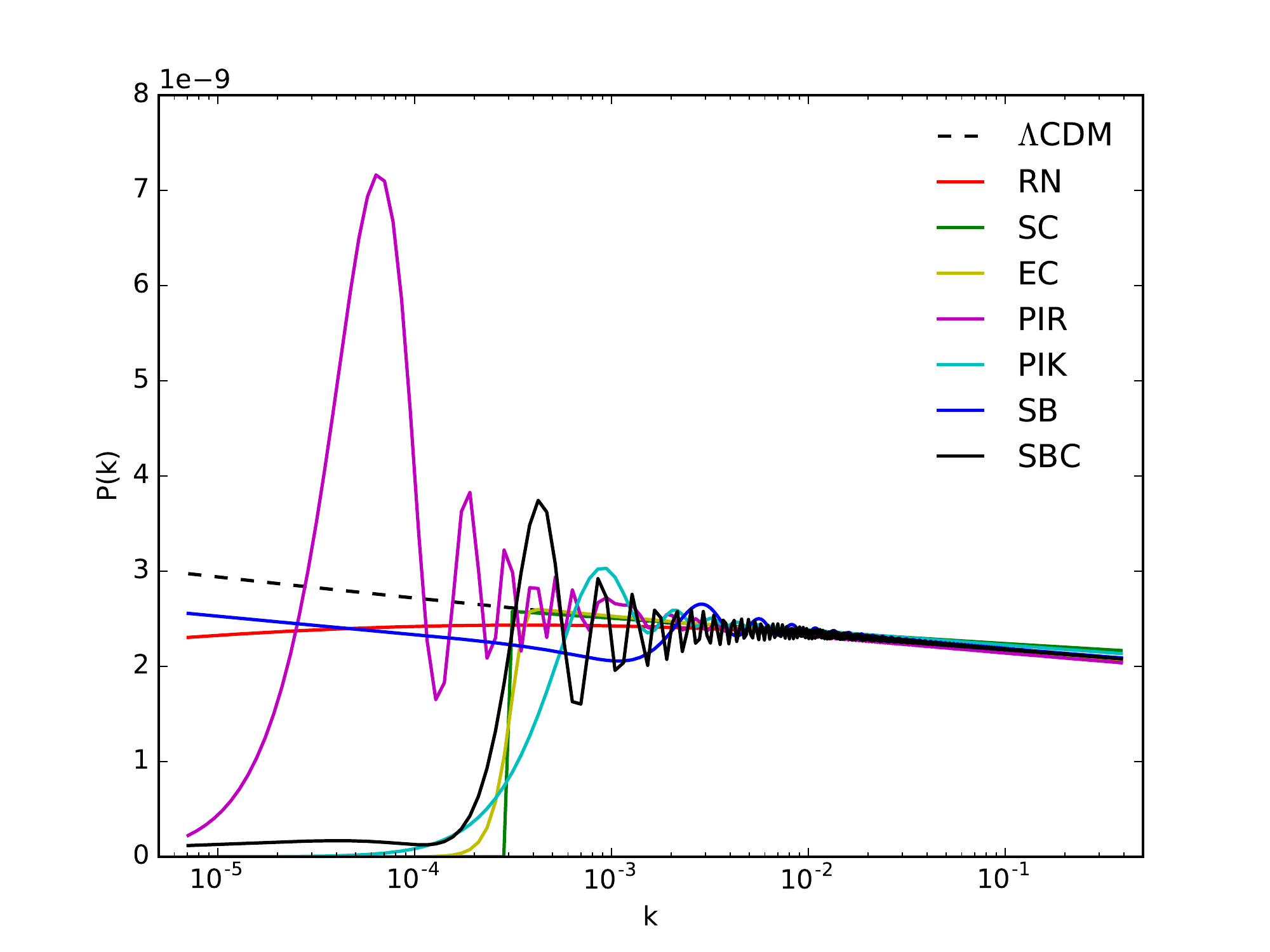}
\end{center}
\caption{Primordial power spectra models considered in this work. The curves use the best-fit parameter values of Tab.~\ref{tab:Tabel_results_2}.}
\label{figprimspec}
\end{figure}
%

Similarly, inflation also predicts tensor perturbations (gravity waves) which produces a tensor spectrum $\mathcal{P}_t(k)$ written as
\begin{eqnarray}
\mathcal{P}_t(k)=A_t\left(\frac{k}{k_0}\right)^{n_t},
\end{eqnarray}
where $A_t$ and $n_t$ are, respectively, the tensor amplitude and tensor spectral index. Again, in terms of the slow-roll parameters we can rewritten them as
\begin{eqnarray}
A_t\simeq\frac{3V(\phi)}{2\pi^2 M_{Pl}^2} \quad \mbox{and} \quad n_t\simeq -\frac{r}{8},
\end{eqnarray}
where $\displaystyle r\equiv \mathcal{P}_t(k)/\mathcal{P_R}(k)\simeq16\epsilon $ is the tensor-to-scalar ratio (the relative amplitude of the tensor to scalar modes). However, considering the recent BICEP2 results~\cite{Ade:2015tva} 
we set $r=0$ or $\mathcal{P}_t(k)=0$.

In the next section, we describe the reference model used in our analysis, the power-law potential, and the inflationary primordial power spectra with infrared cutoff explored in this work. The latter use in general the additional cutoff parameter, $k_c$, which denotes the mode where the model diverges from the reference one. These cutoff potentials can be divided into two main categories, namely, the \textit{empirical parametrizations}, able to produce the low power at high scales, and the \textit{physical motivated} models, that can modulate the primordial potential with an observed feature. For illustration purposes, in Fig.~\eqref{figprimspec} we show the features introduced in the primordial spectrum by such inflationary models using the best fit parameter values of Tab.~\ref{tab:Tabel_results_2}.

\subsection{Power Law (PL)}

The Power Law potential is given by Eq. \eqref{eqPPSPL} and can be considered part of the standard cosmology, as the scalar index $n_s$ and primordial amplitude $A_s$ are included in the minimal set of six cosmological parameters of the $\Lambda$CDM model.  
The most recent temperature data of the Planck Collaboration~\cite{Ade:2015lrj} exclude the exact scale invariance, $n_s = 1$, at more than 5$\sigma$, constraining the spectral index to $n_s = 0.9655 \pm 0.0062$ and the primordial amplitude to $\ln(10^{10}A_s) = 3.089 \pm 0.036$ (we refer the reader to Ref.~\cite{Benetti:2017gvm} for a discussion on the $n_s =1$ case). In our work we choose to use this PL parameterization (dashed black line in Fig.~\ref{figprimspec}) as the reference model. 
It is worth noticing that all the models discussed in this paper can be written as modulation over the power law model, i.e.,
\begin{eqnarray}
P(k)=\mathcal{P}_{PL}(k)\times F(k,\Theta)
\end{eqnarray}
where $F(k, \Theta)$ is the modulation part and $\Theta$ is a vector which characterizes the extra parameters.

\subsection{Running spectral index (RN)}

Possibly, the slightest deviation from the PL power spectrum is obtained by considering the dependence of the spectral index with the scale through the parameter $\alpha_s=dn_s/d\ln{k}$. The ``running of the spectral index"~\cite{kosowsky, easter, kobayashi, czerny, Benetti:2013wla} is the second order deviation from the scale invariance, and can be expressed in terms of slow roll parameters as
\begin{eqnarray}
\alpha_s\simeq -2\xi+16\epsilon\eta-24\epsilon^2.
\end{eqnarray}
Although the variation of the spectral index is expected to be small (of the order of $10^{-3}$ in the slow-roll approximation), this correction leads to a suppression of power at large scale in the power spectrum, as 
showed by the solid red line in Fig.~\ref{figprimspec}.
Even with the low statistical significance wherewith is presently measured\footnote{The running parameter is constraint to $\alpha_s= - 0.0084 \pm 0.0082$ at 68\% CL by the Planck Collaboration using temperature data (Planck TT+lowP)~\cite{Ade:2015lrj}. The joint constraint including high-$\ell$ polarization data is $\alpha_s= - 0.0057 \pm 0.0071$ at 68\% CL.}, its behaviour could point to a deviation from the scale-invariant power law model. It worth mentioning that even though a sizable value of $\alpha_s$ can violate the slow roll approximation, 
there are models in which the running parameter can be large while still respecting the slow-roll approximation~\cite{easter, kobayashi, czerny,  ballesteros1, ballesteros2, Benetti:2016jhf}. 

In the present work we use the standard parametrization for the PL model with running of the spectral index:
%
\begin{equation}
\ln{P(k)}=\ln{A_s} + (n_s-1)\ln{\left(\frac{k}{k_0}\right)}+\frac{\alpha_s}{2}\ln^2{\left(\frac{k}{k_0}\right)}.
\end{equation}
%

\subsection{Sharp cut off (SC)}

The simplest empirical model able to describe the observations at large scales is given by the functional form:
\begin{eqnarray}
P(k)=\left\{\begin{array}{rc}
        A_s\left(\frac{k}{k_c}\right)^{n_s-1},  &\mbox{for} \quad k>k_c\\
        0,  &\mbox{otherwise}
            \end{array}\right. \nonumber
\end{eqnarray}
where, $k_c$, is the scale at which the power drops to zero. Using only one extra parameter, this model is able to produce a spectrum that recover the power law model on small scales, as we can see in Fig. \eqref{figprimspec} (green line). This amounts to saying that the constraints on the cosmological parameters remain unchanged. Notice that we are not concerned about the form of the spectrum near the cutoff, instead we are here interested in parametrising the $k_c$ scale. Previous works have considered this model and found constraints on the cutoff scale $k_c$~\cite{bridges1,bridges2,iqbal2015}.

\subsection{Exponential cut off (EC)}

Another phenomenological parametrization of cutoff can be expressed in the simple exponential form~\cite{cline, contaldi}:
\begin{eqnarray}
P(k)=\mathcal{P}_{PL}(k)\left[1-e^{-(k/k_c)^\alpha}\right],
\end{eqnarray}
where $k_c$ is the scale of the cutoff and $\alpha$ is a measure of its steepness. This model is shown as the yellow curve in Fig.~1.
Like the previous one, it also recovers the simple power law form at small angular scales, such that the constraints on the cosmological parameters are not affected. 

\subsection{Pre-inflationary radiation domination (PIR)}

This model has been proposed in the context of spontaneous symmetry breaking phase transitions~\cite{Vilenkin:1982wt,kolb}, arising in gauge theories of elementary-particle interactions~\cite{kirzhnits}. It considers a Universe containing two components during the phase transition, namely, the radiation and the vacuum energy,  and assumes that the pre-inflationary Universe was in a radiation-dominated phase which eventually  evolved to a vacuum-energy-dominated (or de Sitter) phase~\cite{Vilenkin:1982wt}. As shown in~\cite{Vilenkin:1982wt,powell,wangI}, this can lead to modulations in the PPS, such as an infrared cutoff with a ``bump". 

We consider the following functional form:
\begin{equation}
P(k)=A_sk^{1-n_s}\frac{1}{4y^4}\mid e^{-2iy}(1+2iy)-1-2y^2\mid ^2\;,
\end{equation}
where $y=k/k_c$. 
The cutoff scale $k_c$ is set by the Hubble parameter at the onset of inflation and the current horizon crosses the Hubble radius around the onset of inflation. The behaviour of this potential is shown (magenta line) in Fig. \eqref{figprimspec}.

\subsection{Pre-inflationary kinetic domination (PIK)}

A power spectrum amplitude suppression is also obtained assuming an inflationary stage where the velocity of the scalar field is not negligible, without necessarily meaning the interruption of inflation~\cite{contaldi}. 
However, in order to affect the low-$\ell$ multipoles, this stage should occur very close to the beginning of the 
inflation, e.g., assuming a pre-inflationary phase with domination of the kinetic term. Thus, the difference of the vacuum in the inflationary kinetic domination phase (relative to the fast-rolling inflationary phase) would imprint a feature in the power spectrum at large scales, corresponding to first modes that crossed out of the Hubble radius at the onset of inflation (see \cite{sinhasou} and references therein). 

The PIK model is given by~ \cite{iqbal2015,sinhasou,contaldi}
\begin{eqnarray}
P(k)=\frac{H^2_{inf}}{2\pi^2}k\mid A-B\mid ^2,
\end{eqnarray}
with
%
\begin{eqnarray}
A=\frac{e^{-ik/H_{inf}}}{\sqrt{32H_{inf}/\pi}}\left[\mathcal{H}_0^{(2)}\left(\frac{k}{2H_{inf}}\right)-\left(\frac{H_{inf}}{k}+i\right)\mathcal{H}_1^{(2)}\left(\frac{k}{2H_{inf}}\right)\right]
\nonumber
\end{eqnarray}
\begin{eqnarray}
B=\frac{e^{ik/H_{inf}}}{\sqrt{32H_{inf}/\pi}}\left[\mathcal{H}_0^{(2)}\left(\frac{k}{2H_{inf}}\right)-\left(\frac{H_{inf}}{k}-i\right)\mathcal{H}_1^{(2)}\left(\frac{k}{2H_{inf}}\right)\right],
\nonumber
\end{eqnarray}
where $H_{inf}$ is the Hubble parameter during inflation and $\mathcal{H}_0^{(2)}$ and $\mathcal{H}_1^{(2)}$ stand for the Hankel function of the second kind with order $0$ and $1$, respectively.

Under some conditions~\cite{powell,iqbal2015}, the primordial power spectrum can be rewritten as:
\begin{eqnarray}
P(k)=A'_{s}\left(\frac{k}{k_0}\right)^{n_s-1}\frac{H^2_{inf}}{2\pi^2}k\mid A-B\mid ^2,
\end{eqnarray}
with
\begin{eqnarray}
A_s=A'_s\frac{H^2_{inf}}{2\pi^2}k_0\mid A(k_0)-B(k_0)\mid ^2.
\end{eqnarray}
For our analysis purpose, we will treat the $H_{inf}$ as the extra free parameter, as done in Ref.~\cite{iqbal2015}. This potential is shown in Fig.~\eqref{figprimspec} (light blue line).

\subsection{Starobinsky (SB)}

This model was proposed by Starobinsky~\cite{starobinsky} and assumes that the potential of the effective scalar field, which controls the inflationary phase, has a singularity in the form of a sharp change in its slope. This feature would be able to produce an infrared cut off followed by the bump that arises naturally as the first peak of a damped ringing~\cite{sinhasou}. One can choose the value of the scalar field where the slope changes abruptly to be $\phi_0$, and let the slope of the potential above and below $\phi_0$ be $A_{+}$ and $A_{-}$, respectively, in such way that the general form of the scalar field potential can be expressed as:
\begin{eqnarray}
V(\phi)=\left\{\begin{array}{rc}
        V_0+A_{+}(\phi-\phi_0),  &\mbox{for} \quad \phi>\phi_0\\
        V_0+A_{-}(\phi-\phi_0),  &\mbox{for} \quad \phi<\phi_0
            \end{array}\right. \nonumber
\end{eqnarray}
where $V_0$ is the value of the potential at $\phi=\phi_0$, $A_{+}$ and $A_{-}$ are model parameters greater than $0$. 

The resulting power spectrum for this model is~\cite{starobinsky}
\begin{eqnarray}
P(k)=\mathcal{P}_{PL}(k)\mathcal{D}^2(y,\Delta),
\end{eqnarray}
with $\mathcal{D}^2(y,\Delta)$ being the transfer function responsible for making the underlying power spectrum non-flat around the point $k_c$~\cite{starobinsky,martin,contaldi2}, when $\Delta\phi\approx (\phi-\phi_0)$ is small,
\begin{eqnarray}
\mathcal{D}^2(y,\Delta)=\big[1+\frac{9\Delta^2}{2}\left(\frac{1}{y}+\frac{1}{y^3}\right)^2+\frac{3\Delta}{2}\left(4+3\Delta-\frac{3\Delta}{y^4}\right)\frac{1}{y^2}\cos{2y}+3\Delta\left(1-(1+3\Delta)\frac{1}{y^2}-\frac{3\Delta}{y^4}\right)\frac{1}{y}\sin{2y}\big],
\end{eqnarray}
where $y=k/k_c$ (with $k_c$ denoting the location of the step but without effect on the shape of the spectrum) and $\Delta=\frac{A_{+}-A_{-}}{A_{+}}$. 

In this case we consider the transfer function applied over the simple power law model. It is important to notice that the power spectrum $P(k)$ has a sharp decrease followed by a bump at small $k$ with large oscillations and a flat upper plateau on small scales, for $R=A_{+}/A_{-}<1$ (see \cite{starobinsky,contaldi2} for more details). For $R=A_{+}/A_{-}>1$ the power spectrum $P(k)$ has s step-down like feature (toward large $k$).

\subsection{Starobinsky cut off (SBC)}

The last model we consider is the Starobinsky exponential cutoff~\cite{sinhasou,iqbal2015}:
\begin{eqnarray}
P(k)=\mathcal{P}_{PL}(k)\left[1-e^{-(\epsilon k/k_c)^\alpha}\right]\mathcal{D}^2(y,\Delta),
\end{eqnarray}
where $\mathcal{D}^2(y,\Delta)$ is the transfer function of the Starobinsky feature described in Sec. II.G (SB model) and $\epsilon$ sets the ratio of the two cutoff scales involved.

We follow~\cite{iqbal2015} and fix $\epsilon=1$, which reduces the number of degrees of freedom and the degeneracy problem without affecting the final results. We notice that this model showed the best agreement with previous CMB data when compared with other cutoff potentials~\cite{sinhasou,iqbal2015}.

\section{Methodology}
\label{sec:method}
We adopt a Bayesian approach to model selection, since we are interested in knowing whether the latest cosmological data support the inclusion of extra parameters to explain the features in the primordial power spectrum.  In our analysis, we choose to use the CMB data set from the latest release of the Planck Collaboration~\cite{Aghanim:2015xee}, considering the high-$\textit{l}$ Planck temperature data from the 100-,143-, and 217-GHz half-mission T maps in the range of $30 < \textit{l} <2508$, and the low-P data in the range of $2 < \textit{l} < 29$ by the joint TT,EE,BB and TE likelihood.

The minimal $\Lambda$CDM model with the PL primordial potential is assumed as reference and is parameterized with the usual set of cosmological parameters: the baryon density, $\Omega_{b}h^{2}$, the cold dark matter density, $\Omega_{c}h^{2}$, the ratio between the sound horizon and the angular diameter distance at decoupling, $\theta$, the optical depth, $\tau$, the primordial scalar amplitude, $A_{s}$, and the primordial spectral index $n_{s}$. We also choose to work with very large prior on these parameters, as listed in the first six lines of Tab~\eqref{tab_priors}. 
We consider purely adiabatic initial conditions and fix the sum of neutrino masses to $0.06~eV$, setting the pivot scale at $k_*=0.05$ $\rm{Mpc}^{-1}$. In addition to the parameters above we also vary the nuisance foregrounds parameters~\cite{Aghanim:2015xee}.

The key quantity for Bayesian model comparison is the \textit{Bayesian evidence}, or marginal likelihood, and is calculated here by implementing the nested sampling algorithm of {\sc MultiNest}~\cite{Feroz8,Feroz9}
in the current release of the package {\sc CosmoMC}~\cite{Lewis2}. In order to computing the angular power spectrum of CMB anisotropies for each model considered here, we modify the {\sc CAMB} code~\cite{Lewis00}, included in {\sc CosmoMC}. Finally, we use the  Bound Optimization BY Quadratic Approximation ({\sc BOBYQA}) algorithm~\cite{powell_bobyqa} through the Powell's routines as implemented in {\sc CosmoMC}, to obtain the results for the best-fit values of the parameters and to maximize the likelihood itself.

%
\begin{table}
\centering
\caption{{Priors on the model parameters.}
\label{tab_priors}}
\begin{tabular}{|c|c|c|}
\hline 
Parameter Name & Symbol & Prior Ranges \\ 
\hline 
\hline
Baryon Density & $\Omega_{b}h^{2}$ & $[0.005 : 0.1]$ \\ 
 
Cold Dark Matter Density & $\Omega_{c}h^{2}$ & $0.001 : 0.99]$ \\ 

Angular size of Acoustic Horizon & $\theta$ & $[0.5 : 10.0]$ \\ 
 
Optical Depth & $\tau$ & $[0.01 : 0.8]$ \\ 
 
Scalar Spectral Index & $n_{s}$ & $[0.8 : 1.2]$ \\ 
 
Scalar Amplitude & $\log{10^{10}}A_{s}$ \footnotemark[1]
\footnotetext[1]{$k_0 = 0.05\,\Mpc^{-1}$.} & $[2.0 : 4.0]$ \\ 
 
Hubble Parameter at Inflation & $H_{inf}$ (Mpc$^{-1}$) & $[10^{-7} : 10^{-2}]$ \\ 
 
Running Index & $\alpha_{s}$ & $[-1.0 : 1.0]$ \\ 
 
Cut off Parameter & $k_{c}$(Mpc$^{-1}$) & $[0.0 : 0.01]$ \\ 
 
Cut off Steepness Parameter & $\alpha$ & $[1.0 : 15.0]$ \\ 
 
Starobinsky Parameter & $\Delta$ & $[0.0 : 1.0]$ \\ 
\hline 

\end{tabular} 
\end{table} 

\begin{table}[h]
\centering
\caption{{Revised version of Jeffreys' scale adopted in the analysis~\cite{Trotta1}.}
\label{tab:tab_scale}}
\begin{tabular}{|c|c|}
\hline 
$\ln{B_{ij}}$ & Strength of the evidence \\ 
\hline 
\hline
$0 - 1$ & Inconclusive \\ 
\hline 
$1 - 2.5$ & Weak \\ 
\hline 
$2.5 - 5$ & Moderate \\ 
\hline 
$ > 5$ & Strong \\ 
\hline 
\end{tabular} 
\end{table} 

\begin{table*}
\centering
\caption{{
$68\%$ confidence limits for the cosmological parameters using TT+lowP data.}
\label{tab:Tabel_results_1}}
%
\begin{tabular}{|c|c|c|c|c|c|c|c|}
\hline
&\multicolumn{7}{c|}{Parameter}
\\
\hline
{Model}&
$100\,\Omega_b h^2$& $\Omega_{c} h^2$& $\theta$& $\tau$& $\ln 10^{10}A_s$ \footnotemark[1]
\footnotetext[1]{$k_0 = 0.05\,\Mpc^{-1}$.}& $n_s$& $H_0$
\\
\hline
{\textbf{$\Lambda$CDM}} 	
& $2.222 \pm 0.022$ 
& $0.1197 \pm 0.0021$       
& $1.04085 \pm 0.00045$  
& $0.077 \pm 0.018$ 
& $3.088 \pm 0.034$ 
& $ 0.9654 \pm 0.0059$ 
& $ 67.32 \pm 0.95 $ 
\\
{\textbf{RN}} 
& $2.237 \pm 0.026$
& $0.1196 \pm 0.0021$ 
& $1.04093 \pm 0.00047$ 
& $0.088 \pm 0.021$ 
& $3.112 \pm 0.041$ 
& $0.9652 \pm 0.0062$ 
& $67.51 \pm 0.97$ 
\\
{\textbf{SC}}
& $2.221 \pm 0.023$
& $0.1197 \pm 0.0022$ 
& $1.04086 \pm 0.00048$ 
& $0.082 \pm 0.019$ 
& $3.284 \pm 0.039$ 
& $0.9652 \pm 0.0061$ 
& $67.32 \pm 0.97$ 
\\
{\textbf{EC}}
& $2.224 \pm 0.023$  
& $0.1197 \pm 0.0022$ 
& $1.04087 \pm 0.00048$ 
& $0.084 \pm 0.021$ 
& $3.101 \pm 0.040$ 
& $0.9654 \pm 0.0063$ 
& $67.35 \pm 0.98$ 
\\
{\textbf{PIR}}  
& $2.222 \pm 0.023$ 
& $0.1197 \pm 0.0021$ 
& $1.04086 \pm 0.00046$ 
& $0.076 \pm 0.019$ 
& $2.983 \pm 0.046$ 
& $1.035 \pm 0.0061$ 
& $67.33 \pm 0.94$ 
\\
{\textbf{PIK}}
& $2.225 \pm 0.024$ 
& $0.1194 \pm 0.0023$ 
& $1.04089 \pm 0.00049$ 
& $0.084 \pm 0.021$ 
& $3.100 \pm 0.040$ 
& $0.9662 \pm 0.0064$ 
& $67.45 \pm 1.02$  
\\
{\textbf{SB}}
& $2.222 \pm 0.023$
& $0.120 \pm 0.0022$
& $1.04084 \pm 0.00047$
& $0.085 \pm 0.020$
& $3.104 \pm 0.039$
& $0.9641 \pm 0.0065$
& $67.18 \pm 0.99$
\\
{\textbf{SBC}}
& $2.224 \pm 0.0024$ 
& $0.120 \pm 0.0022$
& $1.04089 \pm 0.00048$
& $0.084 \pm 0.021$
& $3.100 \pm 0.040$
& $0.9659 \pm 0.0064$
& $67.44 \pm 1.00$
\\
\hline
\end{tabular}
\end{table*} 
\begin{table*}
\centering
\caption{{
$68\%$ confidence limits for the primordial parameters using TT+lowP data. 
The $\Delta \chi^2_{best}$ and the $\ln {B}_{ij}$ refers to the difference with respect to the minimal $\Lambda$CDM model.}
\label{tab:Tabel_results_2}}
%
\begin{tabular}{|c c|c|c|c|c|c|c|c|c}
\hline
\multicolumn{2}{|c|}{ ~ } &\multicolumn{7}{c|}{Model}
\\
\hline
\hline
\multicolumn{2}{|c|}{Parameter} & {\textbf{RN}} & {\textbf{SC}} & {\textbf{EC}}  & {\textbf{PIR}} & {\textbf{PIK}} & {\textbf{SB}} & {\textbf{SBC}} \\
\hline   
\multirow{ 2}{*}{$\alpha_{s}$}
& {Mean}
& $ -0.0088 \pm 0.0078$ 
& $ ... $ 
& $ ... $ 
& $ ... $
& $ ... $
& $ ... $
& $ ... $
\\
& {Best-fit}
& $ -0.007 $ 
& $ ... $ 
& $ ... $ 
& $ ... $
& $ ... $
& $ ... $
& $ ... $
\\
\hline
\multirow{ 2}{*}{$10^{4}k_{c}$}
& {Mean}
& $ ... $ 
& $ 2.478 \pm 0.8940 $ 
& $ 2.8131 \pm 1.4872 $ 
& $ <0.5682 $
& $ ... $
& $ <13.66 $
& $ <1.908 $
\\
& {Best-fit}
& $ ... $ 
& $ 3.035  $ 
& $ 3.121 $ 
& $ 0.3434 $
& $ ... $
& $ 8.175 $
& $ 1.288 $
\\
\hline
\multirow{ 2}{*}{$\alpha$}
& {Mean}
& $ ... $ 
& $ ... $ 
& $ <8.499 $ 
& $ ... $
& $ ... $
& $ ... $
& $ 6.983 $ (NL)
\\
& {Best-fit}
& $ ... $ 
& $ ... $ 
& $ 7.170 $ 
& $ ... $
& $ ... $
& $ ... $
& $ 0.3972 $
\\
\hline
\multirow{ 2}{*}{$10^{4}H_{inf}$}
& {Mean}
& $ ... $ 
& $ ... $ 
& $ ... $ 
& $ ... $
& $ 2.475 \pm 1.257 $
& $ ... $
& $ ... $
\\
& {Best-fit}
& $ ... $ 
& $ ... $ 
& $ ... $ 
& $ ... $
& $ 3.475 $
& $ ... $
& $ ... $
\\
\hline
\multirow{ 2}{*}{$\Delta$}
& {Mean}
& $ ... $ 
& $ ... $ 
& $ ... $ 
& $ ... $
& $ ... $
& $ 0.0859 \pm 0.0856$
& $ <0.242 $ 
\\
& {Best-fit}
& $ ... $ 
& $ ... $ 
& $ ... $ 
& $ ... $
& $ ... $
& $ 0.0811 $
& $ 0.621$
\\
\hline
\hline
\multicolumn{2}{|c|}{$\Delta \chi^2_{\rm best}$}  
& $ 0.89  $ 
& $ -1.18 $
& $ 1.94 $
& $ -0.31 $ 
& $ 1.56 $
& $ 3.46 $
& $ -0.21 $
\\
\hline
\multicolumn{2}{|c|}{$\ln \mathit{B}_{ij}$} 
& $ -2.9 $ 
& $ -17.7 $
& $ -6.1 $ 
& $ -19.6 $ 
& $ -19.7 $ 
& $ -14.4 $ 
& $ -12.9 $ 
\\
\hline
\end{tabular}
\end{table*} 

In dealing with model comparison, the evidence is the prime tool to evaluation of a model's performance in the light of the data~\cite{Trotta1}. This quantity is based on Bayes' theorem, given by:
\begin{eqnarray}
p(\theta\mid d,M)=\frac{p(d\mid \theta,M)\pi(\theta\mid M)}{p(d\mid M)},
\end{eqnarray}
which relates the posterior probability for the parameters $\theta$ given the data $d$ under a model $M$, $p(\theta\mid d,M)$, with the likelihood, $p(d\mid \theta,M)$, and the prior probability distribution function (which encodes our state of knowledge before seeing the data), $\pi(\theta\mid M)$. The evidence, the denominator of the Bayes' theorem, is given by
\begin{eqnarray}
p(d\mid M)=\int_{\Omega}{p(d\mid \theta,M)\pi(\theta\mid M)d\theta},
\end{eqnarray}
where $\Omega$ assigns for the parameter space under the model $M$. We use the evidence to discriminate two competing models by taking the ratio
\begin{eqnarray}
B_{ij}\equiv \frac{p(d\mid M_{i})}{p(d\mid M_{j})},
\end{eqnarray}
that is the \textit{Bayes factor} of the model \textit{i} relative to the model \textit{j}. It is worth pointing out that the evidence rewards predictive models~\cite{Liddle_comment}, i.e., models with the ability of make predictions that later turn out to fit the data well. On the other hand, models with a large number of free parameters, not required by the data, are penalised for the wasted parameter space (see Refs.~\cite{Martin:2013nzq, Benetti:2016tvm, Tram:2016rcw, Heavens:2017hkr,  Campista, Graef, Simpson:2017qvj, Benetti:2016ycg, Santos:2016sti} for recent work with Bayesian model selection in cosmology). The most usual way to rank the models of interest is adopting a scale to interpret the values of $\ln{B_{ij}}$ in terms of the strength of the evidence of a chosen reference model $M_{j}$, as showed in Tab~\eqref{tab:tab_scale}. We refer the reader to Ref.~\cite{Trotta1} for a more complete discussion about this scale, which is a revisited and more conservative version of the \textit{Jeffreys' scale}~\cite{jeffrey_scale}.
Finally, we choose to use the most accurate Importance Nested Sampling (INS)~\cite{Feroz13,cameron} to calculate the Bayesian evidence, requiring a INS Global Log-Evidence error of $\leq 0.02$.

\section{Results}
\label{sec:results}

\begin{figure*}
 \begin{center}
 \includegraphics[scale=0.33]{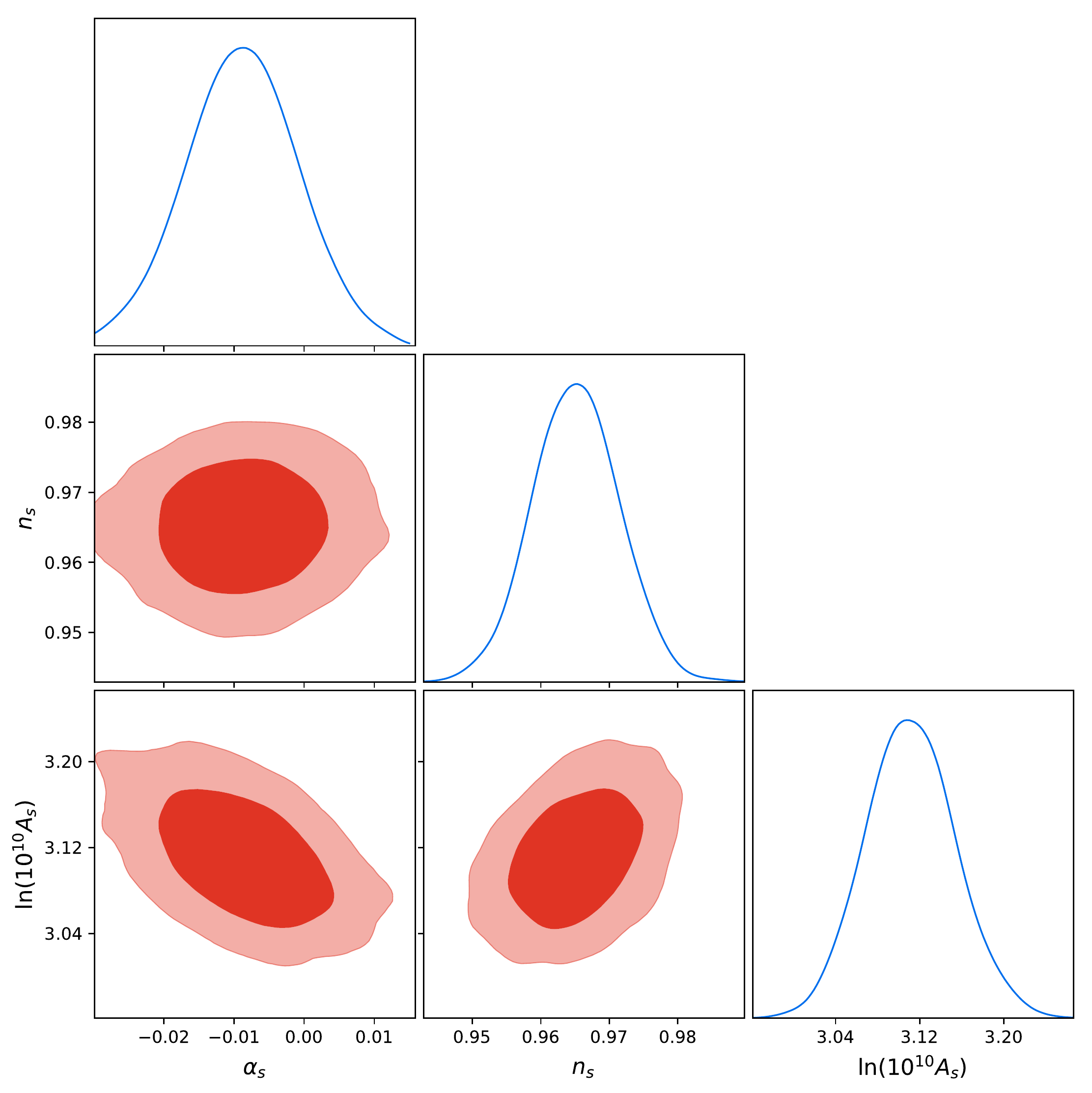}
 \hspace{0.2in}
  \includegraphics[height=3.1in,width=3.6in]{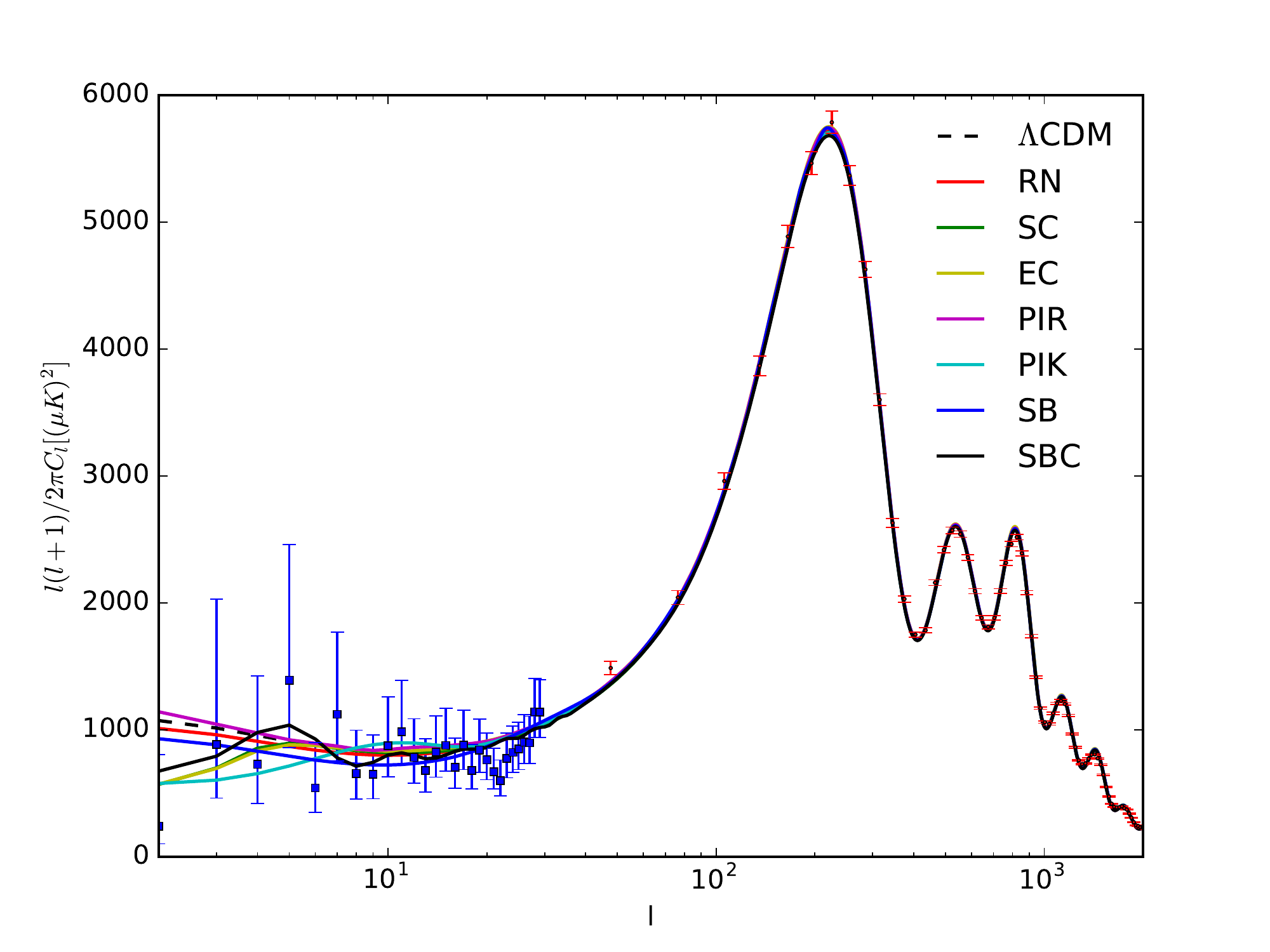}
\end{center}
\caption{{\it{Left:}} Two-dimensional probability distribution and one-dimensional probability distribution for the parameters 
$A_s$, $n_s$ and $\alpha_s$ of the primordial power for the model RN. {\it{Right:}} The best-fit angular power spectra for all models considered in the analysis. The data points correspond to the latest release of Planck data.}
\label{figRNmodel}
\end{figure*}

The main results of our analysis are showed in Tables \ref{tab:Tabel_results_1} and \ref{tab:Tabel_results_2}. There, we list the constraints on the usual cosmological parameters and on the extra primordial parameters for each model considered in this work. In the last lines of the Table \ref{tab:Tabel_results_2}, we present the values of $\Delta\chi^2_{best}$ and $\ln{B_{ij}}$ (Bayes factor), which are obtained considering the $\Lambda$CDM cosmology as the reference model. Fig.~\eqref{figRNmodel} ({\it{left}}) shows the 2D contours and the marginal one-dimensional posterior distributions for the parameters describing the RN model at $68\%$ and $95\%$ confidence level. For brevity, we do not show similar plots for the other models discussed in Sec. II. Additionally, we show in Fig.~\eqref{figRNmodel} ({\it{right}}) the angular power spectra for all models we have examined, which are obtained using the best fit values of Tab.~\eqref{tab:Tabel_results_2}.

We mainly compare our results to the recent work of Iqbal~\textit{et al.} (2015)~\cite{iqbal2015} (IQ15) where a Markov-Chain Monte Carlo analysis was performed using the WMAP-9yr data~\cite{Hinshaw:2012aka} jointly with the first data release of the Planck collaboration (2013)~\cite{Ade:2013kta}. It is worth emphasizing that both the Akaike Information Criteria (AIC) and Bayesian Information Criteria (BIC) adopted by IQ15 differ from the Bayesian model selection approach discussed in Sec. III, as they consider only the point that maximizes the posterior probability distribution to compare the models, taking into account both the number of data points and the number of extra parameters of the models under consideration. 

In what concerns the constraints on the cosmological parameters, we note that the introduction of primordial features does not produce significant changes. The only exception is for the PIR model, whose spectral index $n_s$ mean value grows up until  a red spectral tilt ($n_s>1$). Our results for the RN model are fully consistent with that found by IQ15 and the most recent Plank Collaboration analysis~\cite{Ade:2015lrj}, i.e., confirming that the zero value for the running of the spectral index $\alpha_s$ is off by 68\% (C.L.). Even with a better $\chi^2$ with respect to the PL model, the deviation from the standard cosmological model is too low to be supported from the data and we find that the model is \textit{moderately} discarded, also confirming the recent results of Heavens~\textit{et al.}~\cite{Heavens:2017hkr}. 

For the empirical parameterizations (SC and EC), the constraints for the cut off scale $k_c$ show good agreement with those found by IQ15. However, the Bayesian model comparison performed here shows that the SC model is \textit{strongly} discarted with respect to the PL model, while the AIC value found by IQ15 for this model indicates that it is as good as the PL model. On the other hand, the EC model is disfavoured with respect to the reference model both by Bayesian model comparison as well as by  the AIC and BIC methods.

In what concerns the physical motivated models, the cut off scale of PIR parameterization  is in concordance with the constraints of IQ15. As we can see in the {\it right} pannel Fig.~\eqref{figRNmodel} (magenta line) its prediction is very close to the $\Lambda$CDM curve ($\Delta\chi^2_{best}\sim 0$). For the PIK model, we find a value of $k_c$ bigger than was found by IQ15. This implies an extended oscillation (until $\ell\sim 20$) and a lower power at large scales. The Bayes factor shows that these models are \textit{strongly} disfavoured compared to the PL model, which is in agreement with the IQ15 results.  For the SB model, we find that, although our constraints for the primordial parameters ($k_c$ and $\Delta$) are fully consistent with those found by IQ15, the result of the model selection disagree. The results of  IQ15 (AIC) favour such model with respect to the reference model while our Bayesian analysis indicate that it is \textit{strongly} disfavoured. Finally, the estimates of the SBC primordial parameters ($k_c$, $\Delta$ and $\alpha$) are also in full agreement with those that IQ15 have found, but again according to both Bayesian model comparison and AIC and BIC values this model is \textit{strongly} disfavoured compared to the PL model.

\section{Conclusion}
\label{sec:conclusion}

An intriguing aspect of the current available CMB data is the anomalously low value of the CMB temperature fluctuations up to multipole $\ell < 40$. In the present literature, the most common mechanisms used to address this problem is to consider features in the primordial power spectrum motivated by the early universe physics. 

In this work we have used the most recent data release of the Planck Collaboration and performed a Bayesian model selection analysis to test the observational viability of a set of PPS models. Using a revised version of the Jeffreys' scale given in Ref.~\cite{Trotta1}, our results indicate a \textit{moderate} evidence in favor of the power-law model, adopted in the standard $\Lambda$CDM cosmology, with respect to the Running Spectral Index model (RN), which confirms the recent results of Ref.~\cite{Heavens:2017hkr}.  Moreover, differently from the results obtained in Ref.~\cite{iqbal2015}, we have found that the standard power-law parameterisation is always favoured with \textit{strong} evidence when compared to all the other PPS models considered in our analysis (See Table IV). We believe that the results of the present analysis rule out features in the PPS as a possible explanation for the lack of power observed at large angular scales of the CMB power spectrum and motivate the search for alternative solutions of this open problem.

\section{Acknowledgments}
SSC acknowledges financial support from Coordena\c{c}\~{a}o de Aperfei\c{c}oamento de Pessoal de N\'ivel Superior (CAPES).
MB is supported by the Funda\c{c}\~{a}o Carlos Chagas Filho de Amparo \`{a} Pesquisa do Estado do Rio de Janeiro (FAPERJ - fellowship {\textit{Nota 10}}). JSA is supported by Conselho Nacional de Desenvolvimento Cient\'{\i}fico e Tecnol\'ogico (CNPq) and FAPERJ. 
We also acknowledge the authors of the CosmoMC (A. Lewis) and Multinest (F. Feroz) codes.

%

\end{document}